\documentclass[onecolumn,aps,prl,superscriptaddress]{revtex4-2}
\usepackage[utf8]{inputenc}
\usepackage{epsf,graphicx}
\usepackage{multirow}
\usepackage{amsmath,gensymb,amssymb}
\usepackage{lipsum}
\usepackage{dcolumn}
\usepackage{bm}
\usepackage[hidelinks]{hyperref}
\usepackage{natbib}
\usepackage{color}

\begin{document}

	\title{INCREASING THE RESISTANCE OF MAGNETIC FLUX CONCENTRATOR DURING GENERATION OF STRONG PULSED MAGNETIC FIELDS}

	\author{P. A. Russkikh}
	\email{russkikh\_p@inbox.ru}
	\affiliation{Institute of Electrophysics, Ural Division, Russian Academy of Sciences, Yekaterinburg, 620016 Russia}
		
	\author{G. Sh. Boltachev}
	\affiliation{Institute of Electrophysics, Ural Division, Russian Academy of Sciences, Yekaterinburg, 620016 Russia}

	\begin{abstract}
		The possibility of significant increase of generated pulsed magnetic fields by the inductor system of a single-turn solenoid and magnetic flux concentrator without initialization of low-cycle fatigue mechanism is theoretically studied by varying the size of the inductor system, the material of the concentrator and the parameters of the discharge circuit. The analysis is carried out on the basis of self-consistent solution of the equation, which describes dynamics of the discharge electric circuit, with equations describing spatial distributions of magnetic and temperature fields, mechanical stresses and deformations in the inductor and concentrator. It is shown that for traditionally used steel concentrators by varying the electrical resistance of the circuit it is possible to increase the amplitude of generated pulsed magnetic fields without the threat of concentrator destruction by about 25~\%, from 32 to 40~T.
		
		\end{abstract}

	
	\maketitle

\section{Introduction}

Magnetic-pulse processing is a promising approach for a wide range of technological applications [1-12]. At the same time, the implementation of this approach is hindered by the low lifetime of inductor systems used to generate strong pulsed magnetic fields with an amplitude of about 30-60~T. The low resource is caused by the destruction of the working surface of the conductor limiting the magnetic field region [7-13]. Under the influence of intense thermomechanical stresses accompanying the process of generation of pulsed fields, cracks appear on the surface of the conductor, which rather quickly grow deep due to the "saw effect" [13,14]. The nucleation of initial cracks on the surface of a material having some plasticity resource occurs by the mechanism of low-cycle fatigue [15,16]. In the framework of the elastic–perfectly plastic material, the low-cycle fatigue mechanism is triggered if, during the heating-cooling cycle caused by the flow of surface currents, the conductive material reaches its yield strength twice: during heating and during subsequent cooling [10-12,16]. Therefore, in this study, which is a continuation of [10-12], we will refer to the amplitude of the magnetic induction pulse ${{{B}_{\text{th}}}}$ as the threshold magnetic field, at which plastic flow of the conductive material at the cooling stage begins.

Various approaches [7-12,17-21] have been discussed in the literature to increase the durability of the conductor, in particular: creation of a gradient resistivity profile in the material [7-12,17-19], use of a diamagnetic screen with inertial confinement [20], optimization of the shape of the generated magnetic pulse ${{B}_{0}}(t)$ [17,21]. Under real experimental conditions, the pulse ${{B}_{0}}(t)$ is determined by the dynamics of the discharge circuit and can be represented with sufficient accuracy as a damped sinusoid [7-12]
\begin{equation}
{{B}_{0}}(t)={{B}_{m}}\exp \left( -\frac{t}{{{\tau }_{e}}} \right)\sin \left( \frac{2\pi t}{{{\tau }_{s}}}\right),
\label{Eq1}
\end{equation}
where $\tau_{e}$ is the damping coefficient, $\tau_{s}$ is the oscillation period. In our previous studies [10,11], it has been shown that an appreciable increase in the threshold field ${{{B}_{\text{th}}}}$ can be achieved by decreasing the parameter $\tau_{e}$ or by increasing the period $\tau_{s}$. As noted in [11], the condition $\tau_{s}>\tau_{e}$ corresponds to the effective damping of the electrical oscillations that follow after the first half-wave of the pulse (\ref{Eq1}) and induce excessive heating of the conductor. 

In experiments, the shape of the generated field pulse ${{B}_{0}}(t)$ is depended on the parameters of the discharge circuit, such as its intrinsic resistance ${{R}_{e}}$ and inductance ${{L}_{e}}$. To theoretically analyze the various approaches to increase the lifetime of magnetic flux concentrators and, in particular, the influence of the parameters ${{R}_{e}}$ and ${{L}_{e}}$ on the threshold field ${{{B}_{\text{th}}}}$, the model presented in this paper considers that the inductor, which is a single-turn solenoid with a concentrator placed inside it, is part of a discharge RLC-circuit. Numerical modeling includes self-consistent solution of the equation of dynamics of the circuit with differential equations describing spatial distributions in the solenoid and concentrator of magnetic and temperature fields, mechanical stresses and strains. Achievement of the yield strength of the material is determined due to the Mises yield criterion, and the plastic deformation process is determined in accordance with the associated flow rule [22].

\section{Theoretical Model}
The modeled system consisted of RLC-circuit, solenoid and concentrator is presented in Fig.~1. A single-turn solenoid with inner and outer radii ${{R}_{s,1}}$ and ${{R}_{s,2}}$, respectively, has length ${{L}_{2}}$, which coincides with the length of the outer surface of the concentrator. The amplification of the magnetic field by the concentrator occurs due to decreasing its length $L$ with decreasing radius according to the linear law:
\begin{figure}[h]
	\centering
	\includegraphics[width=0.57\linewidth]{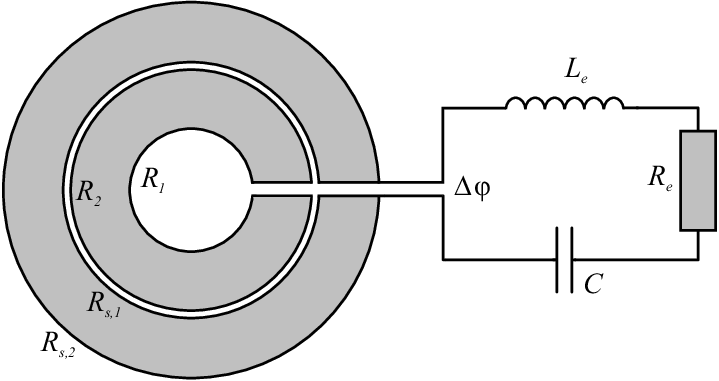}
	\caption{The schematic representation of the modeled system.}
	\label{Fig.1}
\end{figure}
\begin{equation}
	L={{L}_{1}}+{{L}_{r}}(r-{{R}_{1}}),\qquad {{L}_{r}}={({{L}_{2}}-{{L}_{1}})}/{({{R}_{2}}-{{R}_{1}})},
	\label{Eq2}
\end{equation}
where ${{L}_{1}}$ is the length of the inner surface of the concentrator, ${{R}_{1}}$ and ${{R}_{2}}$ are its inner and outer radii, respectively. Since the concentrator is electrically isolated, the total current through it is zero:
\begin{equation}
	\int\limits_{{{R}_{1}}}^{{{R}_{2}}}{j}Ldr=0, \qquad j=-\frac{1}{\mu }\frac{\partial B}{\partial r}.
	\label{Eq3}
\end{equation}
Substituting the expression (\ref{Eq2}) for $L$ into the integral (\ref{Eq3}), and integrating by parts, we get
\begin{equation}
{{B}_{1}}(t)={{B}_{2}}(t)\frac{{{L}_{2}}}{{{L}_{1}}}-\frac{{{L}_{r}}}{{{L}_{1}}}\int\limits_{{{R}_{1}}}^{{{R}_{2}}}{B}dr,\qquad {{B}_{2}}(t)=B({{R}_{s,1}},t)=\frac{\mu I}{{{L}_{2}}},
\label{Eq4}
\end{equation}
where $I$ is the electric current, $\mu$ is the magnetic constant, $B_{1}(t)$ and $B_{2}(t)$ are the magnetic fields at the inner and outer boundaries of the concentrator. The magnetic field inductions at the inner boundary of the solenoid and the outer boundary of the concentrator were assumed to be equal. The expression for $B_{2}(t)$ corresponds to the field of an infinitely long solenoid. To find the current $I$, we consider that both conductors are part of an electric circuit characterized by the intrinsic inductance ${{L}_{e}}$, resistance ${{R}_{e}}$, and capacitance of the capacitor bank $C$ (Fig.~1). The dynamics of the electric circuit is determined by Ohm's law:
\begin{equation}
	U=\Delta \varphi +I{{R}_{e}}+{{L}_{e}}\frac{dI}{dt},\qquad
	I=-\frac{dq}{dt},
	\label{Eq5}
\end{equation}
where $t$ is time, $U=qC$ is the voltage on the capacitor bank, $q$ is its charge, $\Delta\varphi$ is the voltage drop on the solenoid, for which we have
\begin{equation}
	\Delta \varphi ={{j}_{s}}{{\rho }_{s}}2\pi r+\int\limits_{0}^{r}{\frac{\partial B}{\partial t}2\pi rdr}.
	\label{Eq6}
\end{equation}
Here $j_{s}$ is the current density in the solenoid, $\rho_{s}$ is its resistivity. Using the relation (\ref{Eq6}) at radius $r={{R}_{s,1}}$ taking into account expressions (\ref{Eq4}), and substituting it into expression (\ref{Eq5}), we derive the differential equation of the circuit in the form:
\begin{equation}
	\begin{split}
		&{{L}_{eff}}\frac{{{d}^{2}}q}{d{{t}^{2}}}+{{R}_{e}}\frac{dq}{dt}+\frac{q}{C}=\Delta {\varphi }'={{j}_{s,1}}{{\rho }_{s,1}}2\pi {{R}_{s,1}}-\pi R_{1}^{2}\frac{{{L}_{r}}}{{{L}_{1}}}\int\limits_{{{R}_{1}}}^{{{R}_{2}}}{\frac{\partial B}{\partial t}\,dr}
		+2\pi \int\limits_{{{R}_{1}}}^{{{R}_{2}}}{\frac{\partial B}{\partial t}r\,dr},\\
		&{{L}_{eff}}={{L}_{e}}+\pi \mu \left( \frac{R_{s,1}^{2}-R_{2}^{2}}{{{L}_{2}}}+\frac{R_{1}^{2}}{{{L}_{1}}} \right),
	\end{split}
	\label{Eq7}
\end{equation}
where $\Delta{\varphi}'$ is the remaining part of the potential difference on the solenoid after subtracting the terms included in the effective inductance ${{L}_{eff}}$. Initial conditions for the equation (\ref{Eq7}):
\begin{equation}
	q(0)=C{{U}_{0}},\qquad I(0)={{\left. \frac{dq}{dt} \right|}_{t=0}}=0,
\end{equation}
where ${{U}_{0}}$ is the initial charging voltage.

To obtain the numerical solution of the circuit equation (\ref{Eq7}), the spatial distributions of the magnetic field $B(r,t)$ and temperature $T(r,t)$ across the solenoid and concentrator must be found, due to the temperature dependence of the resistivity. These distributions were calculated in one-dimensional, axially symmetric, formulation, i.e., neglecting edge effects at the ends of the inductor system [11,12]. In this case, in the cylindrical coordinate system $(r,\varphi,z)$ with the $Oz$ axis coinciding with the symmetry axis, the magnetic field $\mathbf{B}=\left(0; 0; B(r,t) \right)$, the current density $\mathbf{j}=\left(0; j(r,t); 0 \right)$, the displacement vector $\mathbf{w}=(w(r,t); 0; 0)$, and the main equations -- the magnetic field diffusion equation and the heat conduction equation take the form:
\begin{equation}
		\frac{\partial B}{\partial t}=-\frac{1}{r}\frac{\partial }{\partial r}\left( r\rho j \right),\qquad  j=-\frac{1}{\mu }\frac{\partial B}{\partial r},\qquad 
	    c\frac{\partial T}{\partial t}={{\sigma }_{r}}\frac{\partial {{\varepsilon }_{r}}}{\partial t}+{{\sigma }_{\varphi }}\frac{\partial {{\varepsilon }_{\varphi }}}{\partial t}+\rho {{j}^{2}}+\frac{\lambda }{r}\frac{\partial }{\partial r}\left( r\frac{\partial T}{\partial r} \right).
	\label{Eq9}
\end{equation}
where resistivity $\rho$ is equal to $\rho(r,t)$ in the concentrator or ${{\rho }_{s}}(r,t)$ in the solenoid, $c$ is the volumetric heat capacity, $\lambda $ is the thermal conductivity coefficient, and ${{\sigma }_{i}}$, ${{\varepsilon }_{i}}$, where $(i=r,\varphi,z)$, are the corresponding diagonal elements of stress and strain tensors, and
\begin{equation}
	{{\varepsilon }_{r}}=\frac{\partial w}{\partial r}, \qquad {{\varepsilon }_{\varphi }}=\frac{w}{r},\qquad {{\varepsilon }_{z}}=0.
	\label{Eq10}
\end{equation}
A linear dependence on temperature is assumed for the resistivity:
\begin{equation}
	\frac{{{\rho }_{e}}}{\rho _{e}^{*}}=1+{{k}_{\rho }}(T-{{T}_{0}}),
	\label{Eq11}
\end{equation}
where ${{k}_{\rho}}$ is the temperature coefficient of electrical resistance, $\rho _{e}^{*}$ is the initial resistivity, at temperature $T={{T}_{0}}$.

The initial and boundary conditions for the heat conduction equation (\ref{Eq9}) were given in the form:
\begin{equation}
	T\left( t=0,r \right)={{T}_{0}}, \qquad {{\left. \frac{\partial T}{\partial r} \right|}_{r={{R}_{1}}}}=\frac{{{\alpha }_{q}}}{\lambda }(T-{{T}_{0}}),\qquad
	{{\left. \frac{\partial T}{\partial r} \right|}_{r={{R}_{2}}}}=-\frac{{{\alpha }_{q}}}{\lambda }(T-{{T}_{0}}),
	\label{Eq12}
\end{equation}
where ${{\alpha}_{q}}$ is the heat-transfer coefficient. For the magnetic field diffusion equations in the solenoid and concentrator, the initial conditions were zero, i.e., $B(r,0)\equiv 0$ in the whole inductor system. The boundary conditions for the magnetic diffusion equation (\ref{Eq9}) are defined by equations (\ref{Eq4}) and the equality to zero of the magnetic induction at the outer radius of the solenoid, at $r={{R}_{s,2}}$.

The mechanical problem on the spatial distribution of stresses ${{\sigma}_{i}}$, deformations ${{\varepsilon }_{i}}$, and displacements $w$ was solved only for the concentrator. For the solenoid, the contribution of the corresponding terms to the heat conduction equation was assumed to be zero. The following equations were used as the main equations: the condition of mechanical equilibrium
\begin{equation}
	\frac{\partial {{\sigma }_{r}}}{\partial r}+\frac{{{\sigma }_{r}}-{{\sigma }_{\varphi }}}{r}=\frac{1}{2\mu }\frac{\partial ({{B}^{2}})}{\partial r},
	\label{Eq13}
\end{equation}
with boundary conditions ${{\sigma }_{r}}({{R}_{1}})=0$, ${{\sigma }_{r}}({{R}_{2}})=0$; linear relations between stresses ${{\sigma }_{i}}$ and elastic strains $\varepsilon_{i}^{(e)}$
\begin{equation}
	{{\sigma }_{i}}={{\lambda }_{e}}\left( \varepsilon _{r}^{\left( e \right)}+\varepsilon _{\varphi }^{\left( e \right)}+\varepsilon _{z}^{\left( e \right)} \right)+2{{\mu }_{e}}\varepsilon _{i}^{\left( e \right)}-K{{\alpha }_{V}}(T-{{T}_{0}}),
	\label{Eq14}
\end{equation}
where $(i=r,\varphi ,z)$, ${{\lambda }_{e}}$ and ${{\mu }_{e}}$ are the Lam$\acute{\textnormal{e}}$ coefficients, $K$ is the bulk modulus, ${{\alpha }_{V}}$ is the temperature coefficient of volume expansion. When material deforms purely elastically, i.e., ${{\varepsilon }_{i}}=\varepsilon _{i}^{(e)}$, the presented relations (\ref{Eq10}), (\ref{Eq13}), and (\ref{Eq14}) are sufficient to solve the mechanical problem. The reaching of the elastic-plastic limit was determined by using the von Mises yield criterion [11,12,22]:
\begin{equation}
	{{\left( {{\sigma }_{r}}-{{\sigma }_{\varphi }} \right)}^{2}}+{{\left( {{\sigma }_{r}}-{{\sigma }_{z}} \right)}^{2}}+{{\left( {{\sigma }_{\varphi }}-{{\sigma }_{z}} \right)}^{2}}=2\sigma _{S}^{2}, \qquad
	{{\sigma }_{S}}(T)={{\sigma }_{S,0}}\frac{{{T}_{\text{melt}}}-T}{{{T}_{\text{melt}}}-{{T}_{0}}},
	\label{Eq15}
\end{equation}
where ${{{\sigma}_{S}}}$ is the yield strength of the material in uniaxial tension, ${{T}_{{\text{melt}}}}$ is the melting point. When the limit (\ref{Eq15}) is reached, the total strain ${{{\varepsilon}_{i}}}$ begins to contain elastic $\varepsilon _{i}^{(e)}$ and plastic $\varepsilon_{i}^{(p)}$ parts, i.e., ${{\varepsilon }_{i}}=\varepsilon _{i}^{(e)}+\varepsilon _{i}^{(p)}$, for the unambiguous definition of which additional relations are required:
\begin{align}
		\frac{d\varepsilon _{r}^{(p)}}{dt}\left( {{\sigma }_{\varphi }}-{{\sigma }_{z}} \right)+&\frac{d\varepsilon _{\varphi }^{(p)}}{dt}\left( {{\sigma }_{z}}-{{\sigma }_{r}} \right)+\frac{d\varepsilon _{z}^{(p)}}{dt}\left( {{\sigma }_{r}}-{{\sigma }_{\varphi }} \right)=0,\label{Eq16}\\ 
		&\varepsilon _{r}^{(p)}+\varepsilon _{\varphi }^{(p)}+\varepsilon _{z}^{(p)}=0,\label{Eq17}
\end{align}
which represent the associated flow rule (\ref{Eq16}) and the plastic incompressibility equation (\ref{Eq17}). 

\section{Results and Discussion}
The results of modeling within the framework of the theoretical model presented in the previous section are discussed below. The system of differential equations (\ref{Eq7}), (\ref{Eq9}) and (\ref{Eq13}) was solved numerically. The calculation parameters correspond to the experimental setup of the works [7-9]: ${{R}_{1}}=5$~mm, ${{R}_{2}}=15$~mm, ${{R}_{s,1}}=15.5$~mm, ${{R}_{s,2}}=30$~mm, ${{L}_{1}}=19.25$~mm, ${{L}_{2}}=30$~mm and ${{L}_{2}}=30$~mm and $C=430$ $\mu$F. The material of the solenoid and concentrator is 30KhGSA steel with characteristics: $\rho_{e}^{*}=42\cdot {{10}^{-8}}$ Ohm$\cdot$m, $c=3688$~kJ/(m$^{3}\cdot$K), ${{k}_{\rho }}=1.38\cdot {{10}^{-3}}$~K$^{-1}$, $\lambda =39$~W/(m$\cdot$K), $E=205$~GPa, $\nu =0.3$, ${{\beta }_{V}}=13\cdot {{10}^{-6}}$~K$^{-1}$, ${{\sigma }_{S,0}}=1$~GPa, ${{T}_{\text{melt}}}=1400^\circ$C, ${{{\alpha}_{q}}}=10$~W/(m$^{2}\cdot$K).
			
The electrical parameters of the RLC-circuit, namely, the intrinsic inductance ${{L}_{e,0}}$ and resistance ${{R}_{e,0}}$ were set to match the experimental data on the time dependences of the current in the circuit $I(t)$ and the magnetic field in the inner cavity of the concentrator ${{B}_{1}}(t)$. The obtained values are ${{L}_{e,0}}=27.5$~nH and ${{R}_{e,0}}=1.95$~mOhm. The achieved agreement between theory and experiment at these parameters for capacitor bank charging voltages ${{U}_{0}}=5.2$~kV and 10.0~kV is depicted in Fig.~2. Note that we used only the first three half-periods of the experimental current and magnetic field pulses to reliably determine the parameters ${{L}_{e,0}}$ and ${{R}_{e,0}}$. After the third half-period, the values of current and magnetic field become smaller than the error of their experimental measurement, so in theoretical calculations we assumed $I=0$ after the third half-period.				

\begin{figure}[h]
	\centering
	\includegraphics[width=0.65\linewidth]{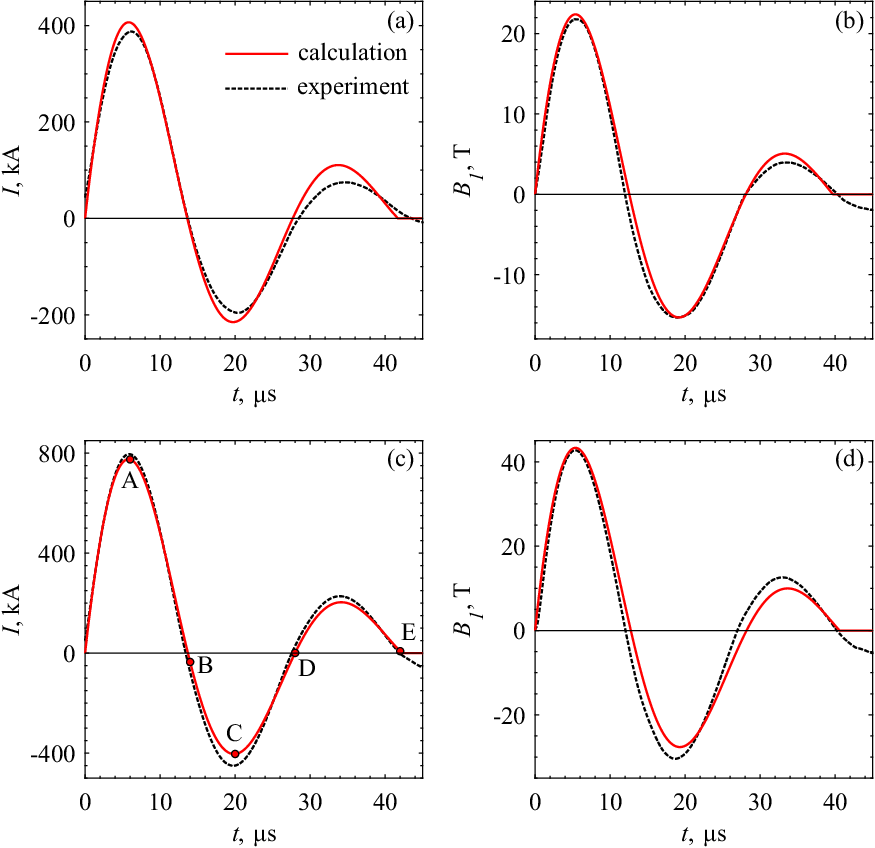}
	\caption{Impulse of current in the electric circuit $I(t)$ (a,c) and magnetic field on the inner surface of the concentrator $B_{1}(t)$ (b,d) at charging voltage $U_{0} = 5.2$~kV (a,b) and 10.0~kV (c,d).}
	\label{Fig.2}
\end{figure}
				
Fig.~3 shows the calculated distributions of the current density across the concentrator and solenoid at a charging voltage ${{U}_{0}}=10$~kV at times $t=6$, 14, 20, 28, and 42 $\mu$s, which correspond to points A-E in Fig.~2. It can be seen that the maximum current density amplitudes are observed on the inner $(r={{R}_{1}})$ and outer $(r={{R}_{2}})$ surfaces of the concentrator, as well as on the solenoid surface at $r={{R}_{s,1}}$. It is these surfaces that are subjected to maximum heating and, as a consequence, maximum thermoelastic stresses. The amplitudes of the current density oscillations decrease rapidly with distance from the surfaces into the depth of the concentrator and solenoid. Note that for a periodic pulse with a duration of ${{{\tau}_{s}}}\cong 28$ $\mu$s, the thickness of a classical skin-layer ${{\delta}_{s}}$ in steel is about 1.7~mm.

\begin{figure}[h]
	\centering
	\includegraphics[width=0.65\linewidth]{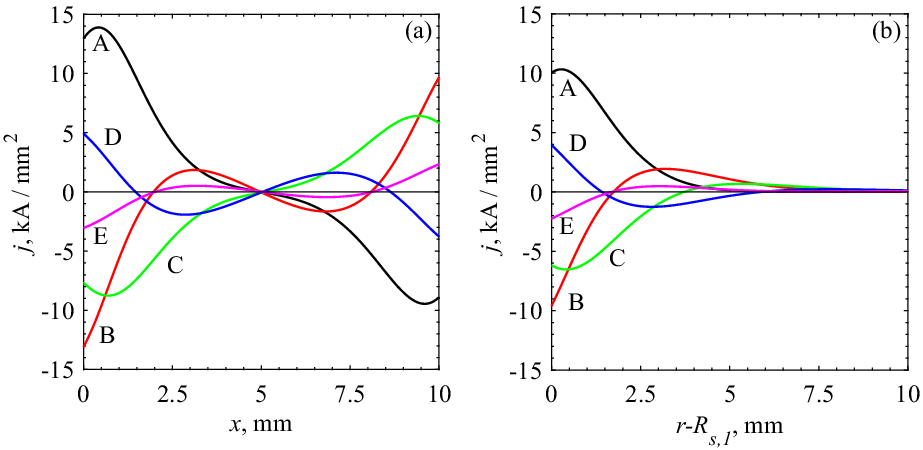}
	\caption{Current density distributions across the concentrator (a) and solenoid (b) at the charging voltage $U_{0}=10$~kV at the moments of time denoted in Fig.~\ref{Fig.2}: $t$ equals to 6 (lines \textit{1}), 14 (\textit{2}), 20 (\textit{3}), 28 (\textit{4}) and 42 $\mu$s (\textit{5}).}
	\label{Fig.3}
\end{figure}
						
The temperatures of the maximum heating of all three surfaces limiting the magnetic field during the discharge of the circuit with the initial charging voltage ${{U}_{0}}=10$~kV are shown in Fig.~4a. The figure shows that the highest heating is observed on the inner surface of the concentrator, where at the moment $t\cong37$ $\mu$s the temperature increment is $\Delta T\cong 865$~K. In comparison, on its outer surface and on the solenoid surface, the maximum heating is approximately 362 and 405~K, respectively. At the heating stage, three consecutive parts corresponding to three half-periods of the current pulse in the discharge circuit are clearly visible. The heating by the end of the first (main) half-period ($t\cong 10$~$\mu$s) is $\Delta T\cong 480$~K, i.e., almost twice lower than the maximum value. In the cooling stage, the initial temperature decrease of all three surfaces ($r={{R}_{1}}$, ${{R}_{2}}$, and ${{R}_{s,1}}$) is mainly due to the heat outflow to the inner less heated layers of the concentrator and solenoid. This process of temperature equalization is completed by $t\cong 1$~s, after which further relatively slow cooling is due to the heat transfer through the surfaces.	

Fig.~4b shows the azimuthal and axial components of the stress tensor on the inner surface of the concentrator. Note that the radial component ${{\sigma}_{r}}$ here is equal to zero according to the boundary condition on the free surface. The figure shows that heating of the material during the discharge leads to a state of tangential (with respect to the surface) compression: ${{{\sigma }_{i}}}<0$ ($i=\varphi,z$). The kink in the curves ${{\sigma }_{i}}(t)$ at the moment $t\cong 3.6$ $\mu$s is associated with the achievement of the yield criterion (\ref{Eq15}) and the beginning of plastic deformation, which unloads the excessive tangential compression. An increase in temperature leads to a decrease in the yield strength ${{\sigma }_{S}}$ due to which, the return of the material to the elastic state, i.e., the cessation of plastic flow, occurs only by the moment $t\cong 45$~$\mu$s. The plastic deformation accumulated in this process leads to irreversibility of the heating-cooling process. As a result, complete cooling of the material to the initial temperature ${{T}_{0}}$ does not lead to the initial state characterized by the absence of stresses. As can be seen in Fig.~4b, at $t>2$~ms the temperature decrease leads to the appearance of positive azimuthal and axial stresses on the surface, i.e., the material is transferred to the state of tangential stretching. The repeated achievement of the yield stress at the moment $t\cong 11$ $\mu$s, which again appears as a kink in the curves ${{\sigma}_{i}}(t)$, guarantees the appearance of plastic deformation at subsequent similar discharges of the capacitor bank, i.e., the initiation of the low-cycle fatigue mechanism.

\begin{figure}[h]
	\centering
	\includegraphics[width=0.65\linewidth]{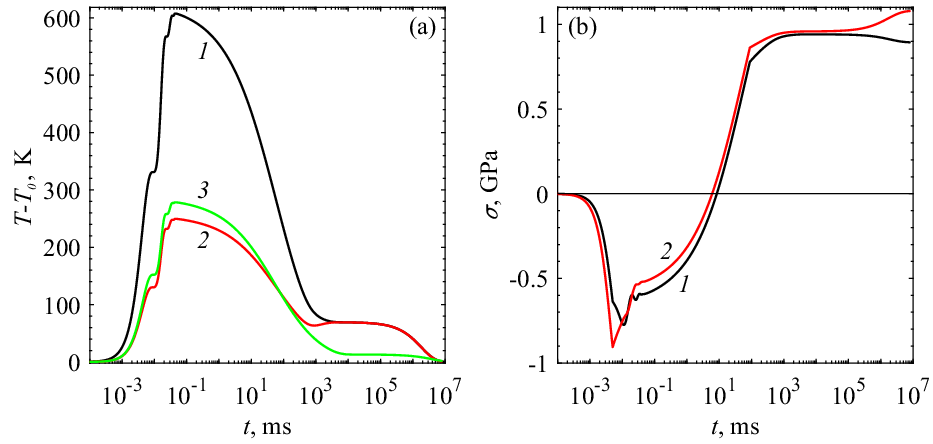}
	\caption{Time dependences of temperature (a) on surfaces $r = R_{1}$ (line \textit{1}), $r = R_{2}$ (line \textit{2}), $r = R_{s,1}$ (line \textit{3}), and stress tensor components $\sigma_{\varphi}$ (line \textit{1}) and $\sigma_{z}$ (line \textit{2}) (b) on surface $r = R_{1}$ at charging voltage $U_{0} = 10$~kV.}
	\label{Fig.4}
\end{figure}

Thus, the maximum amplitude of the magnetic field ${{B}_{\max }}\cong 45.0$~T, corresponding to the discharge with ${{U}_{0}}=10$~kV, exceeds the threshold value ${{{B}_{\text{th}}}}$. Decreasing the charging voltage and, consequently, the amplitude of the magnetic field leads to a shift of the moment of appearance of the second kink in Fig.~4b to the region of larger times and finally to the disappearance, which occurs at a charging voltage ${{{U}_{0}}}=7.06$~kV with a value of ${{{B}_{\text{th}}}}=32.1$~T. Let us analyze how the threshold value of ${{{B}_{\text{th}}}}$ can be increased by varying various parameters of the inductor system. First of all, let us see what is obtained by changing the radii of the concentrator.

\begin{figure}[h]
	\centering
	\includegraphics[width=0.65\linewidth]{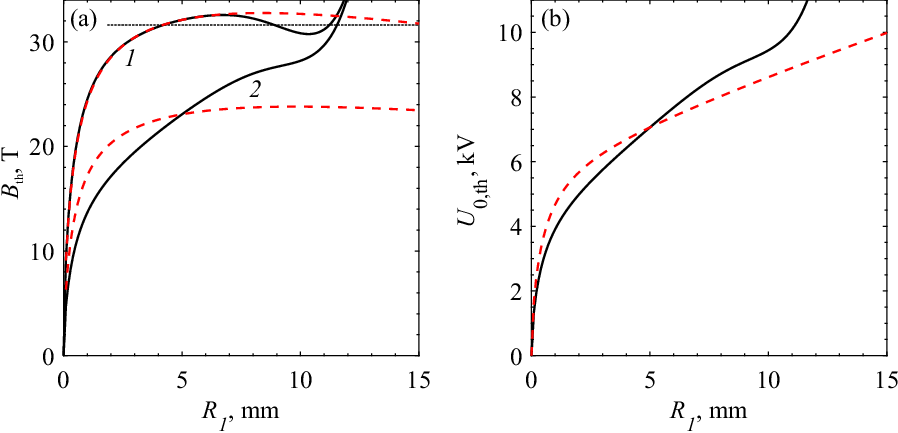}
	\caption{(a) --- threshold fields ${{B}_{\text{th}}}$ (lines \textit{1}), corresponding solenoid fields ${{B}_{2,\max}}$ (lines \textit{2}), and (b) --- charge voltages as a function of the internal radius of the concentrator $R_1$. Dashed red lines: at proportional increase of all radii of the inductor system, i.e., under the condition $R_{i} - R_{1}=\text{const}$; solid black lines - at $R_{i} = {\text{const}}$ ($i = 2$, “$s,1$”, “$s,2$”).}
	\label{Fig.5}
\end{figure}

Fig.~5a shows the calculated threshold fields ${{B}_{\text{th}}}$, which characterize the magnetic fields in the inner cavity of the concentrator, and the field amplitudes generated by solenoid ${{B}_{2,\max}}$, as a function of the inner radius of the concentrator ${{R}_{1}}$. The effect of ${{R}_{1}}$ is analyzed for conditions when all radial dimensions of the inductor system are proportionally increased, i.e., ${{R}_{i}}-{{R}_{1}}=\text{const}$, where $i=2$, “$s,1$” and “$s,2$”, and for conditions when the other dimensions are fixed, i.e., ${{R}_{i}}=\text{const}$. In particular, when all dimensions are proportionally increased to ${{R}_{1}}=15.5$~mm, the threshold field ${{{B}_{\text{th}}}}=31.6$~T (this value is depicted by the horizontal dashed line) corresponds to the limit value of the magnetic field that the solenoid used can withstand without a concentrator. Insertion of the used concentrator (${{R}_{1}}=5$~mm, ${{R}_{2}}=15$~mm) slightly increases the achieved field (${{B}_{\text{th}}}=32.1$~T), but significantly reduces the impact on the solenoid: the field on its surface decreases to the value of ${{B}_{2}}=23$~T. The sharp growth of the threshold field at ${{R}_{1}}>10$~mm for conditions ${{R}_{i}}=\text{const}$ is due to the “degeneration” of the concentrator: as its thickness decreases, at ${{R}_{2}}-{{R}_{1}}\to 0$, first, the distinction between the fields ${{B}_{1}}$ and ${{B}_{2}}$ disappears, and, second, the field begins to penetrate through the concentrator without inducing a current in it. In this case, the concentrator ceases to be subjected to destructive thermal stresses, but also ceases to protect the solenoid: the field on the surface of which at ${{R}_{1}}=11.55$~mm reaches the threshold value ${{B}_{\text{th}}}=31.6$~T. The most optimal values of the inner radius of the concentrator, providing the maximum values of the threshold field, are ${{R}_{1}}=6.7$~mm (i.e., ${{R}_{1}}\approx 4{{{\delta }_{s}}}$) when all other radii are fixed (${{R}_{i}}=\text{const}$) and ${{R}_{1}}=8.4$~mm (i.e., ${{R}_{1}}\approx 5{{{\delta }_{s}}}$) when ${{R}_{i}}-{{R}_{1}}=\text{const}$. The threshold fields realized in this case are ${{B}_{\text{th}}}\cong 32.6$ and 32.8~T, respectively. However, their realization will require a noticeable increase in the charging voltage (see Fig.~5b), up to ${{U}_{0}}\cong 8.1$~kV.			

Another hypothetical way to affect the threshold field ${{{B}_{\text{th}}}}$ is to vary the resistivity of the concentrator material $\rho_{e}^{*}$. Let us analyze the effect of  $\rho_{e}^{*}$ on the field ${{B}_{{\text{th}}}}$, assuming all other system parameters, including the mechanical characteristics of the concentrator, the temperature coefficient of resistance ${{k}_{{\rho }}}$, etc., to be constant. The dependence of ${{B}_{\text{th}}}(\rho _{e}^{*})$ is shown in Fig.~6. As $\rho _{e}^{*}$ decreases from the value corresponding to the steel used $\rho _{\text{st}}^{*}=42\cdot {{10}^{-8}}$~Ohm$\cdot$m, the threshold field increases, up to a value of ${{B}_{\text{th},0}}\cong 40$~T in the hypothetical limit $\rho _{e}^{*}\to 0$, which is approximately 25~\%. The presence of the finite limit ${B}_{{\text{th},0}}$ is due to the fact that at $\rho _{e}^{*}\to 0$ all current flowing through the conducting material is concentrated in the surface layer with a thickness on the order of the skin layer thickness ${{\delta }_{s}}\sim \rho _{e}^{1/2}$. Thus, for a certain value of the surface current, we have a quite certain value of the heat release density:
\begin{equation}
	{{\rho }_{e}}{{j}^{2}}={{\rho }_{e}}{{\left( \frac{I}{L{{\delta }_{s}}} \right)}^{2}}\sim {{I}^{2}}\sim {{B}^{2}},
	\label{Eq18}
\end{equation}
independent of the resistivity. Therefore, the achievement of heating $\Delta T\sim \int{{{\rho }_{e}}{{j}^{2}}dt}$, leading to the appearance of threshold thermal stresses, is determined only by the impulse of the magnetic field $B(t)$, and, in particular, by its amplitude.			

\begin{figure}[h]
	\centering
	\includegraphics[width=0.32\linewidth]{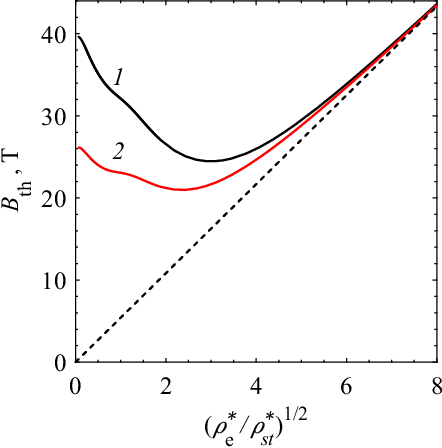}
	\caption{Dependence of the threshold field (line~\textit{1}, ${{B}_{\text{th}}}$) and solenoid field (line~\textit{2}, ${{B}_{2,\max}}$) on the resistivity of the concentrator material. The dashed line shows the asymptotics of ${{B}^2_{\text{th}}}\sim\rho_{e}^{*}$.}
	\label{Fig.6}
\end{figure}

At the opposite limit, at $\rho _{e}^{*}\to \infty$, achieving a certain heat release requires a decrease in the surface current density $j\sim \rho_{e}^{-1/2}$, which according to the diffusion equation (\ref{Eq9}) gives $B\sim\rho _{e}^{*}j\sim \rho _{e}^{1/2}$. As Fig.~6 shows, increasing the resistivity from the value of $\rho _{\text{st}}^{*}$ first decreases the threshold field. At $\rho _{e}^{*}\cong 9\rho _{\text{st}}^{*}$, the dependence of ${{B}_{\text{th}}}(\rho _{e}^{*})$ goes through a minimum. Then, as the resistance is further increased, the threshold field increases, in the limit reaching the asymptotics $B_{\text{th}}^{2}\sim \rho _{e}^{*}$. However, this increase in the “durability” of the concentrator also as in Fig.~5 is related to its “degeneration”: simultaneously with the growth of the field ${{B}_{\text{th}}}$, which the concentrator withstands, the field on the solenoid surface also grows and reaches the threshold value of 31.6~T at $\rho _{e}^{*}\cong 30\rho _{\text{st}}^{*}$.

A promising approach to more significantly increase the threshold field of the inductor system, as shown by the theoretical analysis performed in [10,11], is to optimize the shape of the generated magnetic pulse ${{B}_{2}}(t)$. The increase of ${{{B}_{\text{th}}}}$ is caused by effective damping of electric oscillations following the first half-wave of the pulse, see Fig.~2, and leading to excessive heating of the conductor. The shape of the generated field pulse can be changed by varying such parameters of the discharge circuit as its intrinsic resistance ${{R}_{e}}$ and inductance ${{L}_{e}}$. If it is difficult to decrease these parameters in the experimental setup used, their increase is not complicated. To analyze the influence of the parameters ${{R}_{e}}$ and ${{L}_{e}}$ on the value of the threshold field ${{{B}_{\text{th}}}}$, we estimated the variation of ${{B}_{\text{th}}}$ along the dependences ${{L}_{e}}({{R}_{e}})$ defined by the relation
    \begin{equation}
	{{L}_{e}}({{R}_{e}})={{L}_{e,0}}+\alpha \frac{{{L}_{e,0}}}{{{R}_{e,0}}}({{R}_{e}}-{{R}_{e,0}}),
	\label{Eq19}
	\end{equation}
where $\alpha$ is the proportionality factor. At $\alpha =0$, the increase in resistance ${{R}_{e}}$ is not related to the change in inductance, i.e., ${{L}_{e}}={{L}_{e,0}}$. The dependencies ${{B}_{\text{th}}}({{R}_{e}})$ at values $\alpha$ equal to 0, 0.5, 1.0, and 1.5 are shown in Fig.~7a. We see that increasing the resistance of the RLC-circuit can significantly increase the values of the threshold field, especially if it is not accompanied by an increase in its inductance. Thus, at $\alpha=0$, increasing the resistance to ${{R}_{e}}=9.5{{R}_{e,0}}$ can increase ${{{B}_{\text{th}}}}$ by 26~\%, from 32.1 to 40.3~T. At the same time, the charging voltage increases from ${{U}_{\text{th}}}\cong 7.6$ to 18.8~kV, which is due to a noticeable increase in ohmic losses in the inductor system, and the current pulse transforms into a well-known aperiodic form (see Fig.~7b).

\begin{figure}[h]
	\centering
	\includegraphics[width=0.65\linewidth]{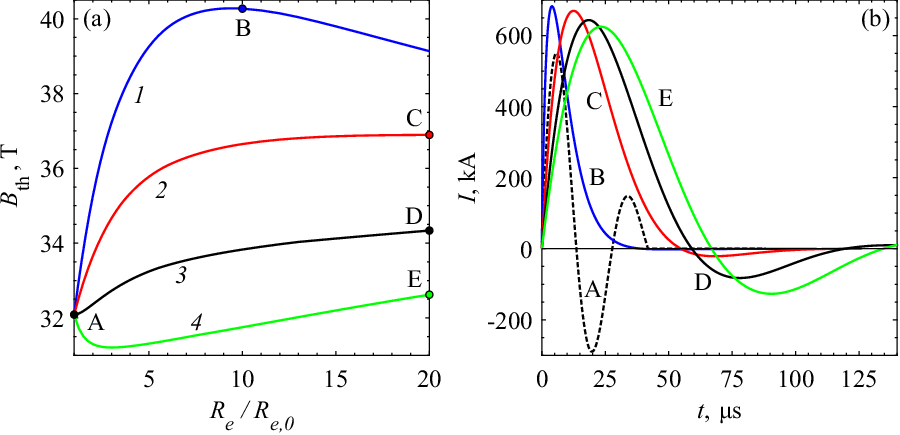}
	\caption{(a) --- dependence of the threshold field ${{B}_{\text{th}}}$ on the electric circuit resistance $R_{e}$ along the dependences $L_{e}(R_{e})$ defined by Eq.~(\ref{Eq19}) with the parameter $\alpha$ equals to 0, 0.5, 1.0 and 1.5. (b) --- time dependences of the current at discharge of the RLC-circuit corresponding to points A-E.}
	\label{Fig.7}
\end{figure}

As the numerical calculations show, along the curve with $\alpha =0.5$ in Fig.~7a, the dependence of ${{B}_{\text{th}}}({{R}_{e}})$ also passes through a maximum at ${{R}_{e}}\cong 21{{R}_{e,0}}$, which, however, is located at a much smaller value, ${{B}_{\text{th}}}\cong 36.9$~T, than the maximum on the curve with $\alpha=0$. The threshold fields ${{{B}_{\text{th}}}}$ at $\alpha$ equals to 1 and 1.5 up to values of ${{R}_{e}}=20{{R}_{e,0}}$ have even smaller values. Fig.~7b shows that on these curves, despite the growth of resistance, the aperiodic shape is not realized, which does not effectively damp the excessive heating of the conducting material at the last stage of magnetic field pulse. Note also that the value of the charging voltage required to generate a magnetic field with threshold amplitude ${{{B}_{\text{th}}}}$ at resistance ${{R}_{e}}=20{{{R}_{e,0}}}$ is ${{U}_{\text{th}}}\cong 40.7$~kV ($\alpha =0.5$), 44.9~kV (1.0), and 48.1 kV (1.5). This significantly exceeds the capabilities of the capacitor bank used in the experimental works [6-9], the maximum charging voltage of which is about 25~kV.
				
Finally, let us analyze the possibility of simultaneous use of high value of the intrinsic resistance ${{R}_{e}}$ and surface modification of the concentrator material, which was investigated in detail in [10-12]. For the modified material of the concentrator, the resistivity instead of Eq.~(\ref{Eq11}) is given by the relations
\begin{equation}
	\frac{{{\rho }_{e}}}{\rho _{e}^{*}}=\gamma (r)+{{k}_{\rho }}(T-{{T}_{0}}),\qquad
	\gamma (r)=1+{{\gamma }_{0}}\exp \left[ -{{\left( \frac{r-{{R}_{1}}}{{{\delta }_{M}}} \right)}^{{{N}_{\gamma }}}} \right],
	\label{Eq20}
\end{equation}
where the function $\gamma (r)$ describes the initial spatial profile of resistance near the working surface of the concentrator; ${{\gamma }_{0}}$ is the “amplitude” of the profile, ${{\delta }_{M}}$ is its characteristic depth. The parameter ${{N}_{\gamma }}$ determines the shape of the initial profile: at ${{N}_{\gamma }}=1$ we have a rather smooth, exponential change in resistivity, and with increasing ${{N}_{\gamma}}$ dependences $\gamma(r)$ become sharper. In accordance with the analysis carried out in [10-12], let us consider modifications with a relatively small amplitude, ${{{\gamma}_{0}}}=1.5$, which can be realized in practice, for example, by methods of ion-plasma treatment [7,8] or diffusion chrome plating [9].

Fig.~8 shows the dependences of the threshold field ${{B}_{{\text{th}}}}$ on the modification depth ${{{\delta}_{M}}}$ at values ${{N}_{\gamma}}$ equal to 1, 2, 6 and for the stepped profile $\left( {{N}_{\gamma }}\to \infty \right)$ for both the initial circuit with the resistance ${{R}_{e,0}}$ and for the circuit with resistance ${{R}_{e}}=10{{R}_{e,0}}$. We see that the increase in the threshold field achieved by increasing the dissipative properties of the RLC-circuit is almost independent of the surface resistivity modifying layers used: the solid and dashed lines in Fig.~8 are almost equidistant. The combined use of the high circuit resistivity ${{R}_{e}}=10{{R}_{e,0}}$ and the most effective modification presented (${{N}_{\gamma }}=2$, ${\delta_{M}}\cong 0.31$~mm) makes it possible to finally increase the limiting amplitude of the pulsed field ${{{B}_{\text{th}}}}$ from 32 to 52~T.

\begin{figure}[h]
	\centering
	\includegraphics[width=0.32\linewidth]{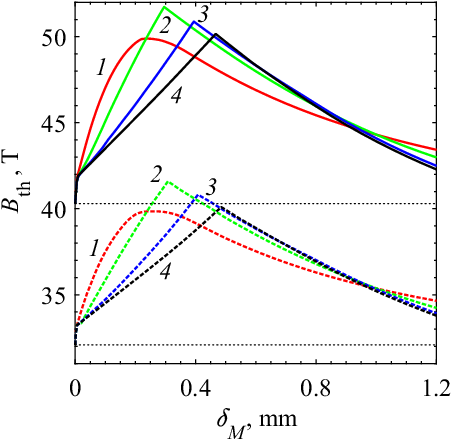}
	\caption{Dependence of the threshold field ${{B}_{{\text{th}}}}$ on the modification depth at “amplitude” $\gamma_{0} = 1.5$ and the parameter $N_{\gamma}$ equals to 1 (lines~\textit{1}), 2 (lines~\textit{2}), 6 (lines~\textit{3}), and $N_\gamma\to\infty$ (lines~\textit{4}). The dashed lines correspond to the initial circuit with resistance $R_{e,0}$; the solid lines correspond to $R_{e} = 10R_{e,0}$.}
	\label{Fig.8}
\end{figure}

\section{Conclusion}
A mathematical model describing the behavior of the inductor system of RLC-circuit, single-turn solenoid and concentrator has built, which considers the dynamics of the discharge electric circuit and the diffusion of magnetic fields both in the solenoid and in the concentrator. Based on the model, a self-consistent solution of the equation of circuit dynamics and equations describing spatial distributions in the inductor and concentrator of magnetic and temperature fields, mechanical stresses and deformations is numerically obtained. The initial values of resistance and inductance of the RLC-circuit $({{R}_{e,0}},{{L}_{e,0}})$ are determined from the condition of the best agreement of the theoretical model with the experimental data of the circuit current and the generated magnetic field. The possibility of a significant increase in the amplitude of ${{{B}_{\text{th}}}}$ of the generated pulsed magnetic fields by the inductor system without initialization of the low-cycle fatigue has been theoretically investigated. The conducted analysis of the influence of the inductor system size has shown that the highest threshold fields ${{{B}_{\text{th}}}}$ are achieved at the inner radius from 4 to 5 skin-layer thicknesses of the steel conductor. Increasing the amplitude of the threshold field is possible either by using materials with higher values of conductivity for the concentrator (with other characteristics unchanged), or by increasing the intrinsic resistance ${{R}_{e}}$ of the electric discharge circuit. Thus, in comparison with the parameters of the experimental setup in [7-9], increasing ${{R}_{e}}$ by a factor of 10 at unchanged intrinsic inductance ${{L}_{e}}$ of the circuit increases the amplitude of the pulsed magnetic fields generated by the used steel concentrators without the threat of their destruction by about 25~\%, from 32 to 40 T. Combined with the use of modifying surface layers with a relatively low amplitude of modification, at which the resistivity ${{\rho}_{e}}$ on the working surface of the concentrator exceeds the resistivity in the material thickness ${\rho_{e}^{*}}$ by 2.5~times, the discovered effect makes it possible to safely achieve even higher magnetic fields~--- about 50~T.

\end{document}